# FABRICATION AND TOLERANCE ISSUES AND THEIR INFLUENCE ON MULTI-BUNCH BBU AND EMITTANCE DILUTION IN THE CONSTRUCTION OF X-BAND RDDS LINACS FOR THE NLC[1]

R.M. Jones, R.H. Miller, T.O. Raubenheimer, and G.V. Stupakov; SLAC, Stanford, CA, USA

*Abstract*

The main linacs of the Next Linear Collider (NLC) will contain several thousand X-band RDDS (Rounded Damped Detuned Structures). The transverse wakefield in the structures is reduced by detuning the modal frequencies such that they destructively interfere and by four damping manifolds per structure which provide weak damping. Errors in the fabrication of the individual cells and in the alignment of the cells will reduce the cancellation of the modes. Here, we calculate the tolerances on random errors in the synchronous frequencies of the cells and the cell-to-cell alignment.

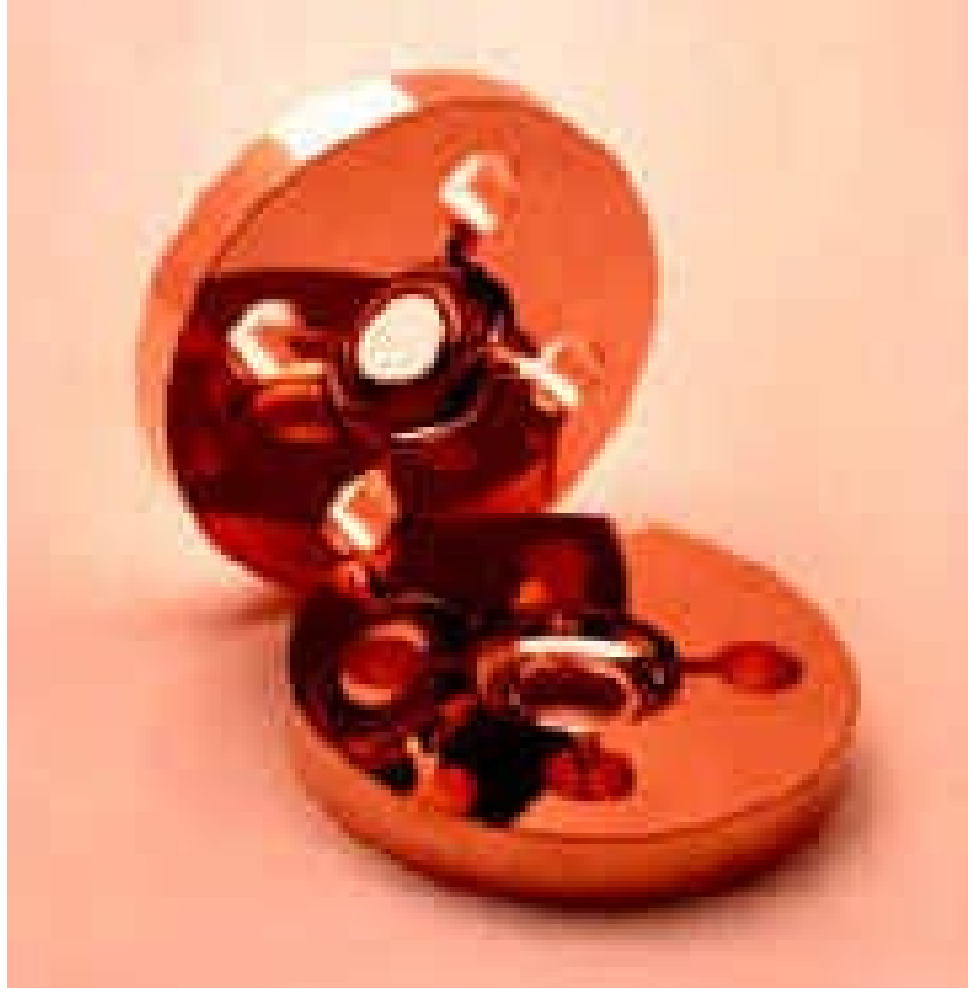

Figure 1: Machined RDDS1 Cells

## 1. INTRODUCTION

In order to answer fundamental questions posed by particle physics a high-energy $e^+$-$e^-$ linear collider is being designed at SLAC and KEK with an initial center-of-mass energy of 500 GeV and the possibility of a later upgrades to 1.0 TeV or 1.5 TeV. The heart of the collider consists of two linear accelerators constructed from approximately 10,000 X-band accelerating structures. These linacs will accelerate a multi-bunch particle beam from 8GeV to 500GeV. Each accelerating structure consists of 206 cells (two of which are shown in Fig. 1) which are bonded together. A displacement of the beam in the structure gives rise to a transverse deflecting force, or wakefield. There are two effects that are of concern: first, the transverse wakefield can cause a multi-bunch beam breakup instability (BBU) which would make the collider inoperable and, second, the wakefields caused by misalignments of the cells and the structures will cause multi-bunch emittance dilution which will reduce the collider luminosity.

The long-range transverse wakefield is reduced by forcing the dipole modes to destructively interfere and damping the modes with four manifolds per structure. However, errors in fabricating and aligning the cells can significantly increase the wakefield and thus it is important to carefully analyse each error component. The following section will discuss the effect of errors in the cell synchronous frequencies and the subsequent section focuses on transverse cell-to-cell and structure-to-structure misalignment errors and the resulting tolerance imposed on the fabrication of the structures for a prescribed multi-bunch emittance dilution.

## 2. MACHINING ERRORS AND EMITTANCE DILUTION

Small dimensional errors, generated when fabricating the irises and cavities of an accelerator structure, give rise to errors in the synchronous frequencies [1]. Presently, it is possible to machine the cells to an accuracy of better than 1 μm [2,3], however, when fabricating several thousand such structures, looser tolerances may reduce the fabrication costs.

The linacs consist of roughly 5000, nominally identical, structures, each of which contains 206 slightly different cells. The nomenclature that we adopt is an error type which is repeated in every cell of a structure but differs in every structure is referred to as: a systematic-random error. Whereas, an error that is repeated in every structure, but varies from cell-to-cell, we refer to as a random-systematic error. We also consider random-random and systematic-systematic (potentially the most damaging) error types making a total of 4 error types. The random errors we consider have an RMS deviation of 3MHz about the mean dipole frequency of the cells. In fabricating RDDS1, the RMS error in the synchronous frequency prior to bonding the cells was 0.5MHz [2,3] and thus simulation of larger errors is pursued with a goal of understanding how much the cell-to-cell fabrication tolerances can be relaxed.

Cell-to-cell frequency errors within an individual structure reduce the effect of the detuning cause a larger wakefield. Although BBU is a complicated effect, an indicator for the onset of BBU is provided by the wakefield at a particular bunch which is formed by summing all wakefields left behind by earlier bunches which is denoted as the “sum wakefield” [4]. BBU will likely arise when the RMS of the sum wake is the order of 1 V/pC/mm/m or larger. When not in the BBU regime, the sum wakefield also provides an accurate method of calculating the multi-bunch emittance dilution and will be used in the following section.

[1] Supported under U.S. DOE contract DE-AC03-76SF00515.

An example of the sum wakefield for a structure with 3MHz RMS errors in the cell synchronous frequencies is plotted in Fig. 2 versus a change in the bunch spacing. Changing the bunch spacing is equivalent to changing all the synchronous frequencies systematically. The wakefield with the random errors is an order of magnitude larger than in a perfect structure and if these cell errors are reproduced in every structure it would be expected to cause significant BBU.

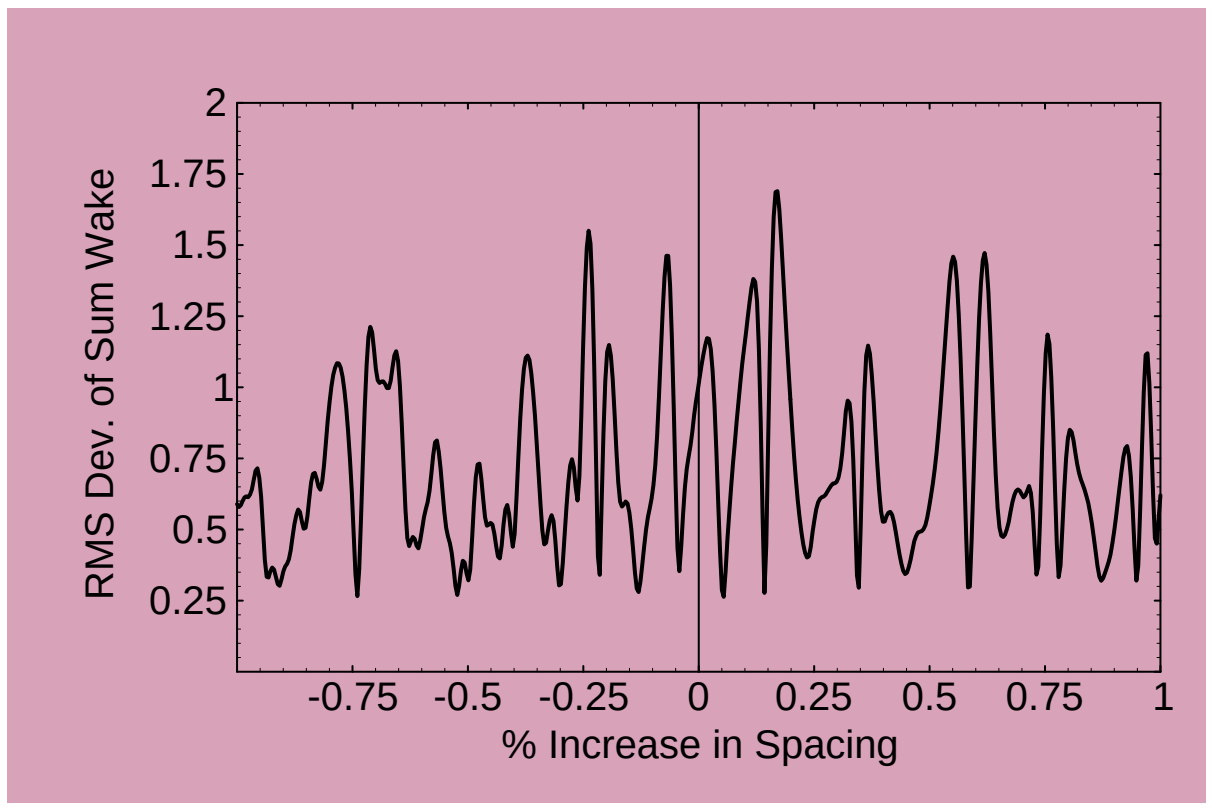

Figure 2. RMS sum wakefield for 3MHz RMS errors

This is confirmed by particle tracking simulations using the code LIAR [5] in which the all structures are assumed to be perfectly aligned and the beam is initially offset by 1μm. When all structures have identical random errors (this is the case of random-systematic errors) and $S_\sigma$ is of the order of unity, the beam clearly undergoes BBU as illustrated in Fig. 3 and the emittance grows by roughly 250%. This is supported by looking at the phase space at the end of the linac which which is plotted in Fig. 4 (a). In contrast, if the cell errors in every structure are different, the random-random case, BBU does not occur and the emittance growth is negligible as is also seen in Fig. 3.

Another important case, is that of an identical systematic error in the synchronous frequencies of the cells and this is investigated by varying the spacing of the bunches in the train of particles. The case of a systematic-systematic error, corresponding to an error in all of the cell frequencies that is repeated in all of the structures, is studied by choosing a particular bunch spacing that results in a peak in the sum wakefield. Such an error also leads to BBU. However, imposing a small random error (3MHz was utilised) from structure-to-structure prevents the resonant growth from occurring; the phase space at the end of the linac corresponding to this random-systematic error is plotted in Fig. 4(b).

The results of relaxing the tolerance are documented in [6] and it is found that even for the very relaxed case of a 5MHz error in the synchronous frequencies BBU does not occur and little emittance growth arises provided this cell-to-cell error is not repeatable over all structures.

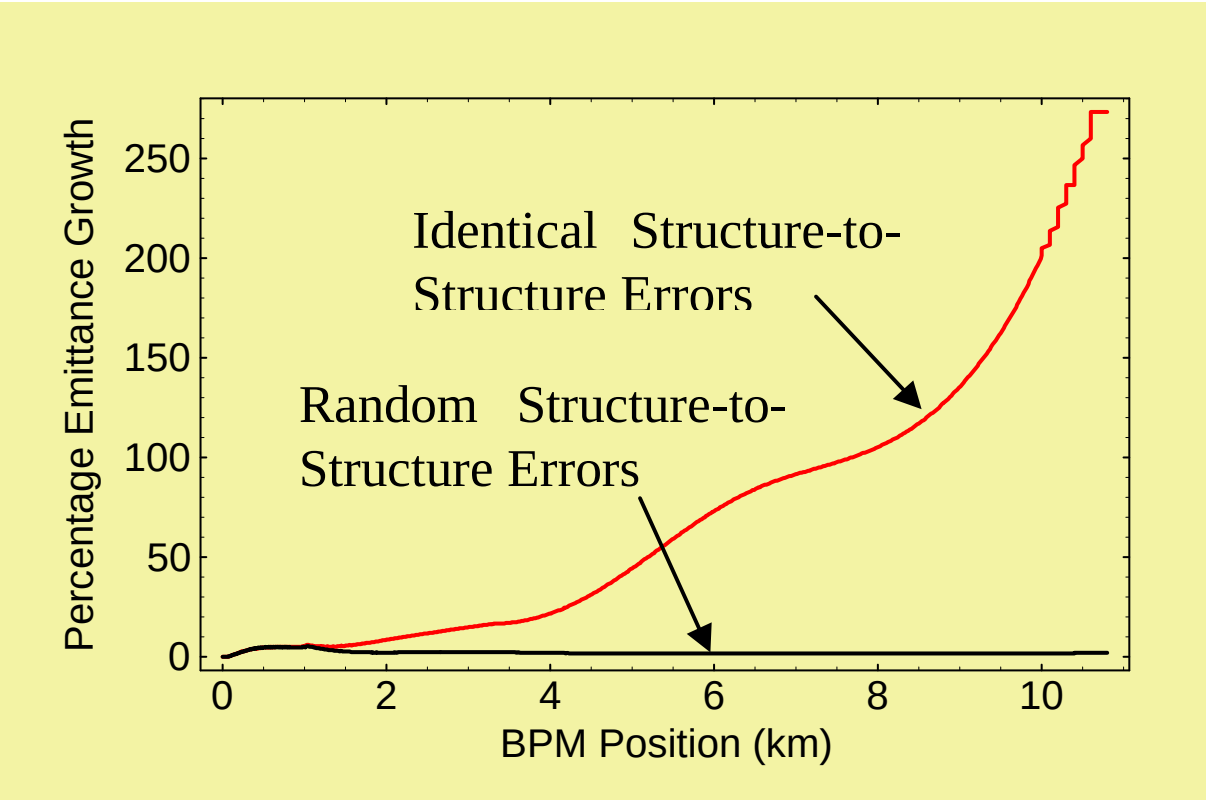

Figure 3. Emittance growth due to 3MHz RMS errors that are (a) reproduced in every structure and (b) random from structure-to-structure.

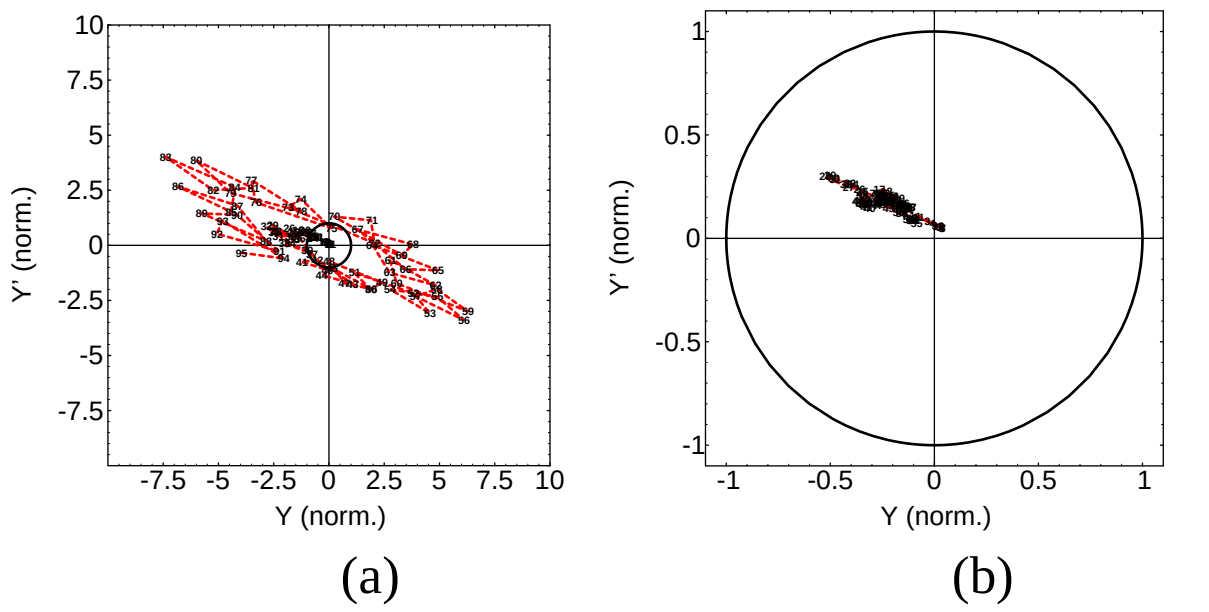

Figure 4. Phase Space (3MHz RMS error). The phase space to the left (a) is for a linac composed of 4720 structures assumed to have identical random errors in each structure. The phase space to the right (b) has been computed from a linac composed of structures with a different random error in the synchronous frequency (non-identical structures).

## 3. TOLERANCES IMPOSED ON STRUCTURE ALIGNMENT

Next, assuming that BBU is not an issue, let us consider the effect of misalignments of the cells and the structures on the multi-bunch beam emittance. In order to estimate the growth of the projected emittance $\Delta\varepsilon$ of a train of bunches caused by misaligned structure cells we uses the following formula for the expectation value of $\Delta\varepsilon$ [5]

$$\langle \Delta\varepsilon \rangle = r_e^2 N^2 \bar{\beta}_0 L_s^2 N_s \langle \Delta S_k^2 \rangle \frac{1-(\gamma_0/\gamma_f)^{1/2}}{\gamma_0^{1/2}\gamma_f^{3/2}} \qquad (3.1)$$

where $r_e$ is the classical electron radius, N is the number of particles in the bunch, $\bar{\beta}_0$ is the average value of the beta function at the beginning of the linac, $N_s$ is the number of structures in the linac, $L_s$ is the length of the structure, $\gamma_0$ and $\gamma_f$ are the initial and final relativistic factors of the beam, and $S_k$ is the sum wake. The quantity $S_k$ is defined as a sum of the transverse wakes $w_i$ generated by all bunches preceding the bunch number k, $S_k = \sum_{i=1}^{k} w_k$ and $\Delta S_k$ is the the difference between $S_k$ and the average value $\langle S \rangle$, with $\langle S \rangle = N_b^{-1} \sum_{k=1}^{N_b} S_k$, where $N_b$ is the number of bunches. Also, $S_\sigma = \langle \Delta S_k^2 \rangle^{1/2}$. Eq. (3.1) is derived assuming a lattice with the beta function smoothly increasing along the linac as $\bar{\beta} \propto E^{1/2}$.

For small misalignments, $w_i$ is a linear function of cell offsets, $w_i = \sum_{k=1}^{N_c} W_{is} y_s$ which can be found from the solution of Maxwell's equations for the structure. The matrix $W_{is}$ for the NLC structure RDDS1 with 206 cells is based on the method described in [7]. It has a dimension of $N_b$ x 206. In our calculation we used $N_b$=95 for bunch spacing 2.8 ns.

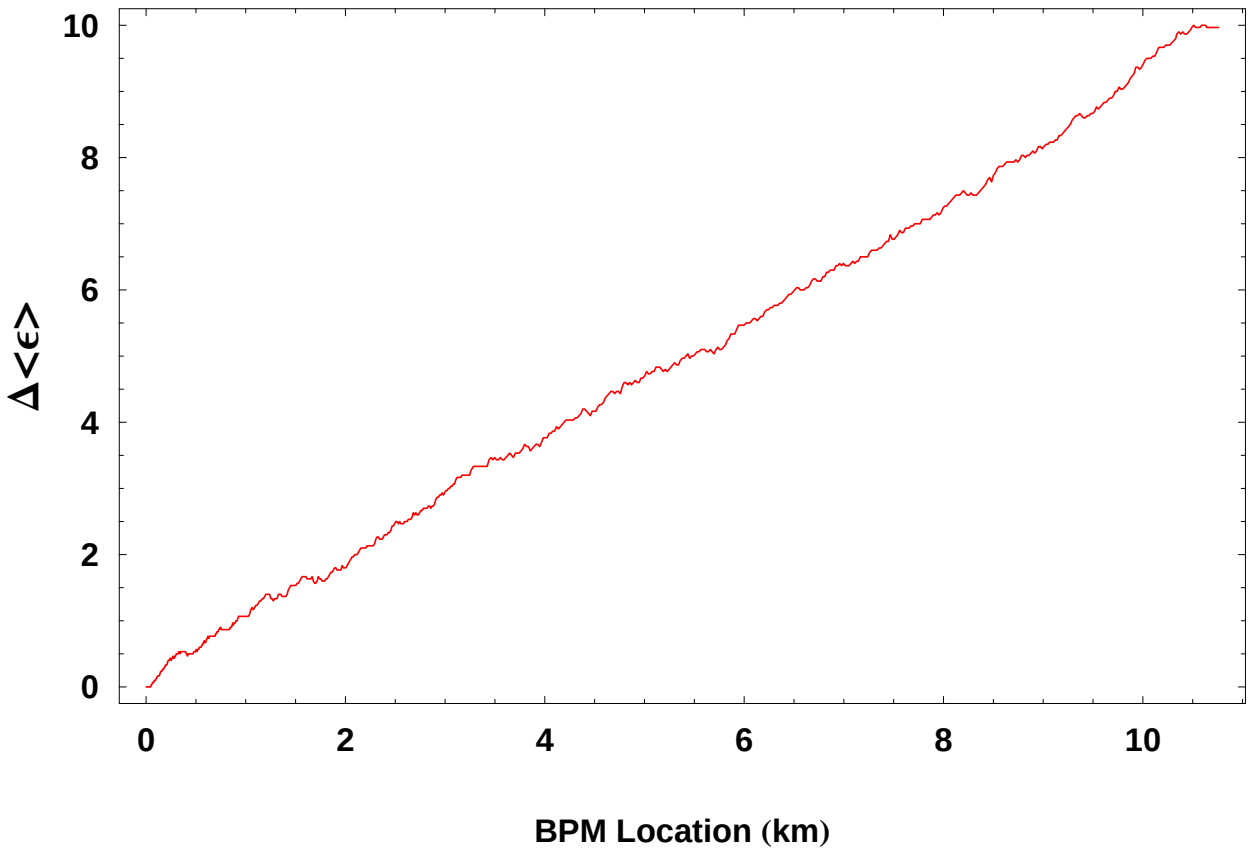

Figure 5: Percentage emittance growth down the linac calculated with the tracking code LIAR for complete structures which are individually offset in a random manner. The RMS offset of the structures is 40µm.

In order to verify Eq. (3.1), we tracked a multi-bunch beam through the complete linac (approx 11km) using the computer code LIAR [4] for RDDS1 with the following linac parameters: beam final energy - $E_f$=500 GeV, number of structures in the linac – $N_s$ = 4720, and number of particles in the bunch – N =1.1 x$10^{10}$. The multi-bunch emittance growth which arises with structures rigidly misaligned is shown in Fig. 5. It is evident that structures randomly misaligned with an RMS value of 40µm gives rise to an overall emittance growth of 10% which grows linearly along the linac as predicted by Eq. (3.1).

The result of many simulations in which each structure is divided up into groups of cells and each individual group is moved randomly transverse to the axis of the linac is illustrated in Fig. 6. It is seen that the analytical formula based on the sum wakefield (line) generally agrees well with the LIAR simulation (points). It should be noted that the single bunch emittance growth due to rigid structure misalignments imposes a much more severe tolerance than that due to the multi-bunch emittance growth [8] however the multi-bunch effects sets that tolerances on the alignment of the individual cells and short pieces of the structure. The tolerance for the cell alignment is about 6µm in this piecewise model. Alternately, assuming a random walk model for the cell-to-cell alignment [9], each cell must be aligned with respect to its neighbour with an RMS of 2~3µm.

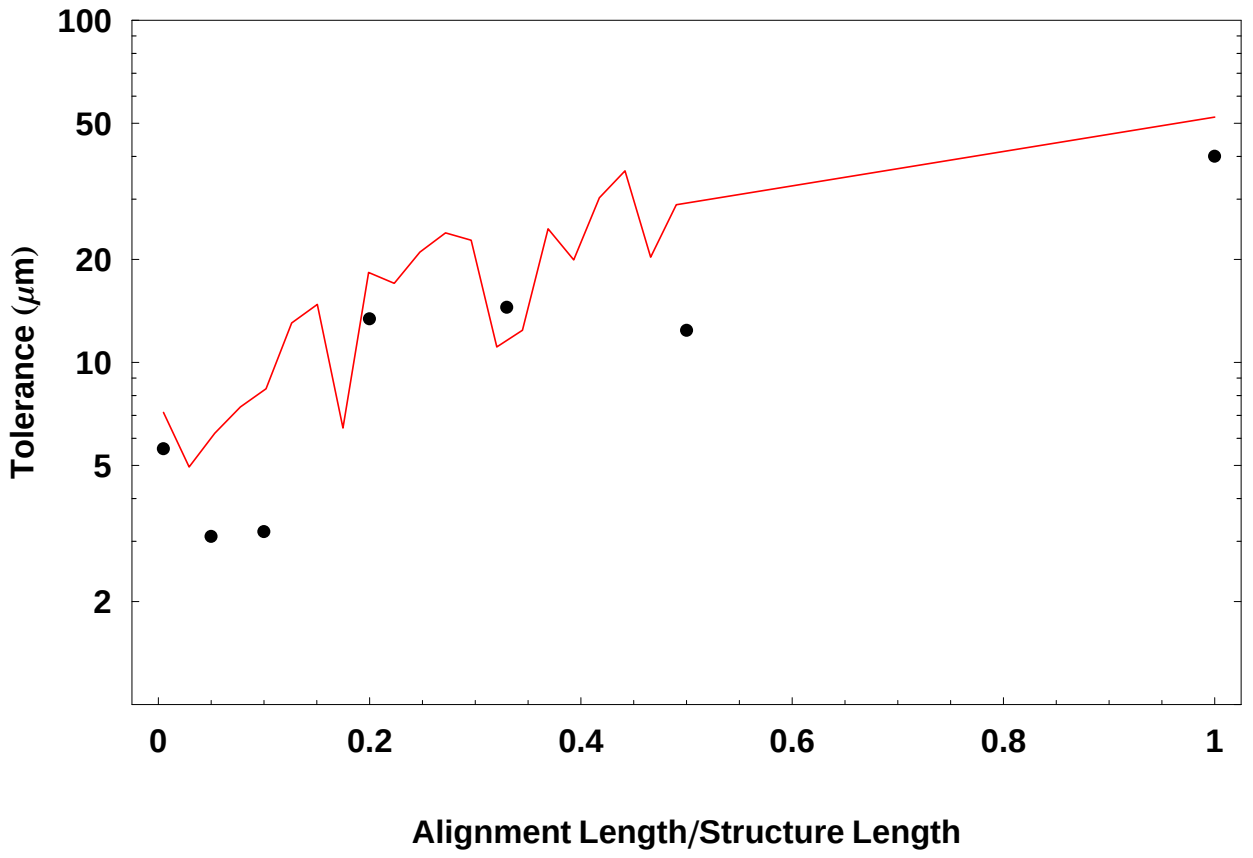

Figure 6: Tolerance vs. misalignment length in units of the structure length $L_s$ for 10% multi-bunch emittance dilution. The solid curve shows the result of the analytical calculation based on Eq. (3.1); dots are the tolerances calculated using LIAR

## 4. CONCLUSIONS

We have discussed four distributions of frequency errors. BBU will arise in the NLC from cell frequency errors of many MHz which are repeated in every structure. However, in practise it is expected that fabrication errors will occur randomly from cell-to-cell *and from structure-to-structure* and hence BBU is unlikely to occur. Furthermore, to meet a prescribed multi-bunch emittance growth of 10%, the cells in the present RRDS structure design structure must be aligned to better than 6µm and the average alignment of the structure must be better than 40µm. Of course, the average alignment tolerance is dominated by single bunch tolerances and must be closer to 10µm[8].